\documentclass[twocolappendix,appendixfloats,]{emulateapj}

\usepackage[colorlinks = true, linkcolor = blue, urlcolor  = blue, citecolor = blue, anchorcolor = blue]{hyperref}
\usepackage{hyperref}
\usepackage{apjfonts}
\usepackage{rotating}
\usepackage{amsmath}
\usepackage{color}
\usepackage{graphicx}
\usepackage[dvipsnames]{xcolor}
\usepackage{CJK}

\newcommand{\muvec}{\mbox{\boldmath $\mu$}}

\newcommand{\thetae}{\theta_{\rm E}}
\newcommand{\pie}{\pi_{\rm E}}

\definecolor{darkbrown}{RGB}{139,69,19}

\shorttitle{OGLE-2018-BLG-0740L\lowercase{b}}
\shortauthors{Han et al.}

\begin{document}

\title{Spectroscopic Mass and Host-star Metallicity Measurements for Newly Discovered Microlensing
Planet OGLE-2018-BLG-0740L\lowercase{b}}


\author{
Cheongho~Han\altaffilmark{0001}, 
Jennifer~C.~Yee\altaffilmark{0002,201},
Andrzej~Udalski\altaffilmark{0003,202},
Ian~A.~Bond\altaffilmark{0004,203},
Valerio~Bozza\altaffilmark{0005,0006},
Arnaud~Cassan\altaffilmark{0007},
Yuki~Hirao\altaffilmark{0008,203}, 
Subo~Dong\altaffilmark{0009},
Juna~A.~Kollmeier\altaffilmark{0010},
Nidia~Morrell\altaffilmark{0011}, 
Konstantina~Boutsia\altaffilmark{0011},\\
(Leading authors),\\
Michael~D.~Albrow\altaffilmark{0012}, 
Sun-Ju~Chung\altaffilmark{0013,0014},  
Andrew~Gould\altaffilmark{0013,0015,0016},
Kyu-Ha~Hwang\altaffilmark{0013}, 
Chung-Uk~Lee\altaffilmark{0013}, 
Yoon-Hyun~Ryu\altaffilmark{0013},
In-Gu~Shin\altaffilmark{0013}, 
Yossi~Shvartzvald\altaffilmark{0017}, 
Youn~Kil~Jung\altaffilmark{0013},
Doeon~Kim\altaffilmark{0001}, 
Woong-Tae~Kim\altaffilmark{0018}
Sang-Mok~Cha\altaffilmark{0013,0019}, 
Dong-Jin~Kim\altaffilmark{0013}, 
Hyoun-Woo~Kim\altaffilmark{0013}, 
Kyeongsoo~Hong\altaffilmark{0001},
Seung-Lee~Kim\altaffilmark{0013,0014}, 
Dong-Joo~Lee\altaffilmark{0013}, 
Yongseok~Lee\altaffilmark{0013,0019}, 
Byeong-Gon~Park\altaffilmark{0013,0014}, 
Richard~W.~Pogge\altaffilmark{0015}, 
Weicheng~Zang\altaffilmark{0020}\\
(The KMTNet Collaboration),\\
Przemek~Mr{\'o}z\altaffilmark{0003},
Micha{\l}~K.~Szyma{\'n}ski\altaffilmark{0003},
Jan~Skowron\altaffilmark{0003},
Radek~Poleski\altaffilmark{0015},
Igor~Soszy{\'n}ski\altaffilmark{0003},
Pawe{\l}~Pietrukowicz\altaffilmark{0003},
Szymon~Koz{\l}owski\altaffilmark{0003},
Krzysztof~Ulaczyk\altaffilmark{0021},
Krzysztof~A.~Rybicki\altaffilmark{0003},
Patryk~Iwanek\altaffilmark{0003},
Marcin~Wrona\altaffilmark{0003}\\
(The OGLE Collaboration) \\   
Fumio~Abe\altaffilmark{0022}, 
Richard~Barry\altaffilmark{0023},           
David~P.~Bennett\altaffilmark{0023,0024},
Aparna~Bhattacharya\altaffilmark{0023,0024}, 
Martin~Donachie\altaffilmark{0025}, 
Akihiko~Fukui\altaffilmark{0026,0027},  
Yoshitaka~Itow\altaffilmark{0022}, 
Kohei~Kawasaki\altaffilmark{0008},              
Iona~Kondo\altaffilmark{0008}, 
Naoki~Koshimoto\altaffilmark{0028,0029}, 
Man~Cheung~Alex~Li\altaffilmark{0025},    
Yutaka~Matsubara\altaffilmark{0022}, 
Yasushi~Muraki\altaffilmark{0022}, 
Shota~Miyazaki\altaffilmark{0008},                
Masayuki~Nagakane\altaffilmark{0008}, 
Cl\'ement~Ranc\altaffilmark{0023}, 
Nicholas~J.~Rattenbury\altaffilmark{0025}, 
Haruno~Suematsu\altaffilmark{0008}, 
Denis~J.~Sullivan\altaffilmark{0030}, 
Takahiro~Sumi\altaffilmark{0008},                
Daisuke~Suzuki\altaffilmark{0031}, 
Paul~J.~Tristram\altaffilmark{0032}, 
Atsunori~Yonehara\altaffilmark{0033}\\                    
(The MOA Collaboration)\\
}

\email{cheongho@astroph.chungbuk.ac.kr}


\altaffiltext{0001}{Department of Physics, Chungbuk National University, Cheongju 28644, Republic of Korea} 
\altaffiltext{0002}{Center for Astrophysics | Harvard \& Smithsonian, 60 Garden St., Cambridge, MA 02138, USA}
\altaffiltext{0003}{Warsaw University Observatory, Al.~Ujazdowskie 4, 00-478 Warszawa, Poland} 
\altaffiltext{0004}{Institute of Natural and Mathematical Sciences, Massey University, Auckland 0745, New Zealand}
\altaffiltext{0005}{Dipartimento di Fisica ``E.~R.~Caianiello'', Universit\'e di Salerno, Via Giovanni Paolo II, I-84084 Fisciano (SA), Italy}
\altaffiltext{0006}{Istituto Nazionale di Fisica Nucleare, Sezione di Napoli, Via Cintia, I-80126 Napoli, Italy}
\altaffiltext{0007}{Institut d'Astrophysique de Paris, Sorbonne Universit\'e, CNRS, UMR 7095, 98 bis boulevard Arago, 75014 Paris, France}
\altaffiltext{0008}{Department of Earth and Space Science, Graduate School of Science, Osaka University, Toyonaka, Osaka 560-0043, Japan}
\altaffiltext{0009}{Kavli Institute for Astronomy and Astrophysics, Peking University, Yi He Yuan Road 5, Hai Dian District, Beijing 100871, China}
\altaffiltext{0010}{Observatories of the Carnegie Institution of Washington, 813 Santa Barbara St., Pasadena, CA 91101, USA}
\altaffiltext{0011}{Las Campanas Observatory, Carnegie Observatories, Casilla 601, La Serena, Chile}
\altaffiltext{0012}{University of Canterbury, Department of Physics and Astronomy, Private Bag 4800, Christchurch 8020, New Zealand} 
\altaffiltext{0013}{Korea Astronomy and Space Science Institute, Daejon 34055, Republic of Korea} 
\altaffiltext{0014}{Korea University of Science and Technology, Korea, (UST), 217 Gajeong-ro, Yuseong-gu, Daejeon, 34113, Republic of Korea}
\altaffiltext{0015}{Department of Astronomy, Ohio State University, 140 W.\ 18th Ave., Columbus, OH 43210, USA} 
\altaffiltext{0016}{Max Planck Institute for Astronomy, K\"onigstuhl 17, D-69117 Heidelberg, Germany} 
\altaffiltext{0017}{IPAC, Mail Code 100-22, Caltech, 1200 E.\ California Blvd., Pasadena, CA 91125, USA}
\altaffiltext{0018}{Department of Physics \& Astronomy, Seoul National University, Seoul 151-742, Republic of Korea}
\altaffiltext{0019}{School of Space Research, Kyung Hee University, Yongin, Kyeonggi 17104, Korea} 
\altaffiltext{0020}{Physics Department and Tsinghua Centre for Astrophysics, Tsinghua University, Beijing 100084, China} 
\altaffiltext{0021}{Department of Physics, University of Warwick, Gibbet Hill Road, Coventry, CV4 7AL, UK} 
\altaffiltext{0022}{Institute for Space-Earth Environmental Research, Nagoya University, Nagoya 464-8601, Japan}
\altaffiltext{0023}{Code 667, NASA Goddard Space Flight Center, Greenbelt, MD 20771, USA}
\altaffiltext{0024}{Department of Astronomy, University of Maryland, College Park, MD 20742, USA}
\altaffiltext{0025}{Department of Physics, University of Auckland, Private Bag 92019, Auckland, New Zealand}
\altaffiltext{0026}{Instituto de Astrof\'isica de Canarias, V\'ia L\'actea s/n, E-38205 La Laguna, Tenerife, Spain}
\altaffiltext{0027}{Department of Earth and Planetary Science, Graduate School of Science, The University of Tokyo, 7-3-1 Hongo, Bunkyo-ku, Tokyo 113-0033, Japan}
\altaffiltext{0028}{Department of Astronomy, Graduate School of Science, The University of Tokyo, 7-3-1 Hongo, Bunkyo-ku, Tokyo 113-0033, Japan}
\altaffiltext{0029}{National Astronomical Observatory of Japan, 2-21-1 Osawa, Mitaka, Tokyo 181-8588, Japan}
\altaffiltext{0030}{School of Chemical and Physical Sciences, Victoria University, Wellington, New Zealand}
\altaffiltext{0031}{Institute of Space and Astronautical Science, Japan Aerospace Exploration Agency, 3-1-1 Yoshinodai, Chuo, Sagamihara, Kanagawa, 252-5210, Japan}
\altaffiltext{0032}{University of Canterbury Mt.\ John Observatory, P.O. Box 56, Lake Tekapo 8770, New Zealand}
\altaffiltext{0033}{Department of Physics, Faculty of Science, Kyoto Sangyo University, 603-8555 Kyoto, Japan}

\altaffiltext{201}{KMTNet Collaboration.}
\altaffiltext{202}{OGLE Collaboration.}
\altaffiltext{203}{MOA Collaboration.}

\begin{abstract}
We report the discovery of the microlensing planet OGLE-2018-BLG-0740Lb.  The 
planet is detected with a very strong signal of $\Delta\chi^2\sim 4630$, but 
the interpretation of the signal suffers from two types of degeneracies. One type 
is caused by the previously known close/wide degeneracy, and the other is caused 
by an ambiguity between two solutions, in which one solution requires to 
incorporate finite-source effects, while the other solution is consistent with 
a point-source interpretation.  Although difficult to be firmly resolved based 
on only the photometric data, the degeneracy is resolved in strong favor of the 
point-source solution with the additional external information obtained from astrometric 
and spectroscopic observations.  The small astrometric offset between the source and 
baseline object supports that the blend is the lens and this interpretation 
is further secured by the consistency of the spectroscopic distance estimate of the 
blend with the lensing parameters of the point-source solution.  The estimated mass 
of the host is $1.0\pm 0.1~M_\odot$ and the mass of the planet is $4.5\pm 0.6~M_{\rm J}$ 
(close solution) or $4.8\pm 0.6~M_{\rm J}$ (wide solution) and the lens is located at 
a distance of $3.2\pm 0.5$~kpc.  The bright nature of the lens, with $I\sim 17.1$ 
($V\sim 18.2$), combined with its dominance of the observed flux suggest that 
radial-velocity (RV) follow-up observations of the lens can be done using 
high-resolution spectrometers mounted on large telescopes, e.g., VLT/ESPRESSO, 
and this can potentially not only measure the period and eccentricity of the 
planet but also probe for close-in planets.  We estimate that the expected RV 
amplitude would be $\sim 60\sin i ~{\rm m~s}^{-1}$.
\end{abstract}


\section{Introduction}\label{sec:one}

Microlensing provides a tool to detect exoplanets because a planetary companion
to a lens can manifest its presence through the perturbation to the lensing light curve
produced by the host of the planet \citep{Mao1991, Gould1992b}. The
characteristics of the planetary signal varies depending on the lens-system configurations,
and the analysis of the signal enables one to determine the planet/host mass ratio,
$q$, and the projected planet-host separation in units of the angular Einstein radius 
$\thetae$, $s$.
However, determining these planet parameters is often hampered by various types of
degeneracy, which lead to multiple interpretations of the observed signal. Finding the
types of degeneracies and understanding their origins are important to identify similar
degeneracies in subsequent analyses and thus to correctly interpret the observed signal.

The types and origins of degeneracies for some specific cases of planetary signals 
are known. The ``close/wide degeneracy'' is the most well-known type, which causes 
difficulty in distinguishing the perturbations produced by central caustics induced
by planetary companions with separations $s$ and $s^{-1}$. This degeneracy is intrinsic in the
sense that it is rooted in the symmetry of the lens equations between the lenses with $s$
and $s^{-1}$ \citep{Griest1998, Dominik1999, An2005}.  The ``binary-source/planet
degeneracy'' is an accidental degeneracy, which causes difficulty in distinguishing a
short-term planetary anomaly from the anomaly produced by a subset of binary-source
events with a small flux ratio between the binary-source stars and the close approach of
the faint source companion to the lens \citep{Gaudi1998}. It was recently found that this
degeneracy not only applies to a short-term anomaly but also can extend to various
cases of planetary lens system configurations \citep{Jung2017, Shin2019, Dominik2019}.

With the increasing number of planetary microlensing events, various types of degeneracies 
have been newly identified.  Many of these degeneracies are caused by the ambiguity in determining 
the exact source trajectory with respect to the caustic. Such an ambiguity was first 
predicted by \citet{Gaudi1997}, who pointed out that the magnification pattern on the near 
and far sides of the major-image caustic, which represented the planetary caustic produced 
by a planet with $s>1.0$, were similar, and thus the anomalies produced by the source 
approaching both sides of the caustic were similar to each other: ``major-image degeneracy''. 
\citet{Han2018a} pointed out that for some specific lens-system configurations, a similar 
degeneracy could occur for planetary anomalies produced by the minor-image caustic, which 
was produced by a planet with $s<1.0$.  \citet{Skowron2018} found that a major-image degeneracy 
could also occur in the case of anomalies resulting from the source star's caustic crossings. 
This ``caustic-chiral degeneracy'', which occur when there is a gap in data, results in 
similar values $s$ but substantially different $q$, while the degeneracy between two 
non-caustic-crossing degenerate solutions considered by \citet{Gaudi1997} results in 
similar value of $q$.  \citet{Hwang2018} reported a new type of discrete degeneracy 
between the solution in which the major-image caustic was fully enveloped and the 
solution in which only one side of the caustic was enveloped.  The two solutions subject to 
this so-called ``Hollywood degeneracy''  \citep{Gould1997} result in different mass ratios 
because the source passes through the caustic in different places relative to its center.

In this paper, we present the analysis of the microlensing event OGLE-2018-BLG-0740,
which exhibits a strong short-term anomaly produced by a planetary companion.  
We find that the interpretation of the planetary signal suffers from a new type of 
discrete degeneracy caused by the incomplete coverage of the planetary anomaly.  In 
section~\ref{sec:two}, we mention the data acquisition and processing.  In 
section~\ref{sec:three}, we describe the procedure of the data analysis and depict 
the degeneracy found from the analysis.  We characterize the source star in 
section~\ref{sec:four} and present the physical parameters of the planetary system 
estimated from Bayesian analysis in section~\ref{sec:five}.  In section~\ref{sec:six}, 
we present the external information that enables to resolve the degeneracy.  
In section~\ref{sec:seven}, we discuss the possibility of further characterizing the 
planetary system, including measuring the planet's period and eccentricity as well 
as probing for additional planets, using radial velocity (RV) measurements.
In section~\ref{sec:eight}, we summarize the results and conclude.

\section{Observation and Data}\label{sec:two}

The source star of the microlensing event OGLE-2018-BLG-0740 is located toward 
the Galactic bulge field with equatorial coordinates 
$({\rm RA}, {\rm decl.})_{\rm J2000}= (18:08:42.47, -29:50:08.9)$, which correspond 
to the Galactic coordinates $(l,b)=(1.74^\circ, -4.80^\circ)$. The apparent baseline 
brightness of the star before lensing magnification was $I_{\rm base}\sim 16.87$. 
We note that the source is heavily blended as we will show in section~\ref{sec:four} 
and only $\sim 2\%$ of the measured flux comes from the source star.

The lensing event was first found by the Optical Gravitational Lensing Experiment 
\citep[OGLE:][]{Udalski2015} on 2018-05-08 
(${\rm HJD}^\prime\equiv {\rm HJD}-2450000\sim 8246$) 
when the source had apparently brightened by $\sim 0.14$ magnitude from the baseline, 
and the discovery of the event was notified to the microlensing community. 
On 2018-05-16 (${\rm HJD}^\prime\sim 8254$), 
the event was also found by the Microlensing Observations in Astrophysics 
(MOA) group \citep{Bond2001, Sumi2003}. In the ``MOA Transient Alerts'' page, 
the event was listed as MOA-2018-BLG-147. OGLE observations were conducted mostly 
in $I$ band, with occasional $V$-band observations for the source-color measurement, 
with $\sim 1$ day cadence using the 1.3~m telescope located at Las Campanas 
Observatory in Chile. MOA observations were carried out in a customized broad 
$R$ band with $\sim 1$ hr cadence using the 1.8~m telescope located at the Mt.~John 
University Observatory in New Zealand.

The event was independently discovered by the Korea Microlensing Telescope Network 
\citep[KMTNet:][]{Kim2016} survey in its annual post-season analysis \citep{Kim2018} 
and was designated as KMT-2018-BLG-1822. KMTNet observations were conducted using 
three identical 1.6~m telescopes that are located at the Siding Spring Observatory, 
Australia (KMTA), Cerro Tololo Interamerican Observatory, Chile (KMTC), and the 
South African Astronomical Observatory, South Africa (KMTS). During the period near 
the anomaly, KMTNet observed this field with cadence of 2.5 hours from KMTC and 3.3 
hours from KMTA and KMTS.  KMTNet observations were conducted both in $I$ and $V$ 
bands, and 1/10 of KMT $I$-band observations are complemented by $V$-band images.

\begin{figure}
\includegraphics[width=\columnwidth]{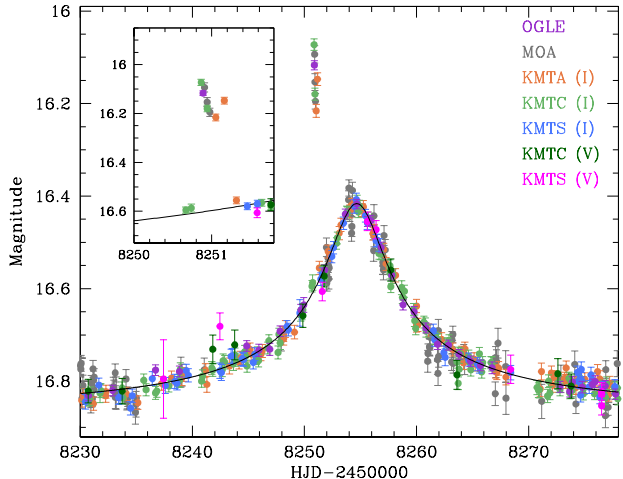}
\caption{
Light curve of the microlensing event OGLE-2018-BLG-0740. 
The solid curve superposed on the data points represents the 
single-lens and  single-source 
(1L1S) model obtained by fitting the data excluding 
those around the anomaly at ${\rm HJD}^\prime\sim 8251$.
The inset shows the enlarged view of the anomaly.
\smallskip
}
\label{fig:one}
\end{figure}

In addition to the usual photometric observations, additional spectroscopic follow-up 
observations were conducted using  
the Inamori Magellan Areal Camera and Spectrograph \citep[IMACS,][]{Dressler2011}
mounted on the 6.5~m Magellan-Baade telescope at Las Campanas Observatory in 
Chile.  As will be discussed in section \ref{sec:four}, the spectroscopic observations 
were conducted to specify the stellar type of the blend object.

The event was analyzed in real time with its progress.  On 2018-05-14 
(${\rm HJD}^\prime\sim 8252$), V.~Bozza noticed a single point anomaly at 
${\rm HJD}^\prime=8250.89$ in the OGLE data, which was confirmed to be real by the 
OGLE group.  The anomaly was additionally confirmed in real-time by the MOA (3
points) and later by KMTNet data sets (3 KMTC and 2 KMTA points).  Since the anomaly 
was confirmed, a series of models describing the anomaly were circulated by 
V.~Bozza, A.~Cassan, and Y.~Hirao.  Although there existed slight variation in 
detailed lensing parameters, all models agreed that the anomaly was produced by a 
planetary companion.

Photometric data sets used in the analysis are processed using the photometry 
codes of the individual survey groups: \citet{Udalski2003}, \citet{Bond2001}, 
\citet{Albrow2009} for the OGLE, MOA, and KMTNet surveys, respectively.  All 
of these codes are based on the difference imaging technique developed by 
\citet{Alard1998}.  We normalize the error bars of the individual data sets 
following the procedure described in \citet{Yee2012}.  In order to measure 
the source color, we additionally conduct photometry using the pyDIA photometry 
\citep{Albrow2017} for a subset of the KMTNet data (KMTC $I$- and $V$-band 
data sets).

In Figure~\ref{fig:one}, we present the light curve of the lensing event.  The 
curve superposed on the observed data points represents the model based on the 
single-lens (1L) and single-source (1S) modeling  excluding the anomaly part of 
the data.  The inset shows the enlarged view of the anomaly, which occurred at 
$t_{\rm anom} \sim 8251$. The duration of the anomaly, which lasted less 
than a day, is short.  Apart from the anomaly, one finds that the event is 
well described by a 1L1S model.

\section{Interpretation of the Anomaly}\label{sec:three}

For the interpretation of the anomaly, we conduct modeling of the light 
curve. The observed short-term anomaly is a characteristic feature produced 
by a planetary companion to the lens, and thus we first conduct modeling under 
the assumption that the lens is composed of two masses: 2L1S model. Because 
it is known that such an anomaly could in principle also be produced
by a companion to a source, we also conduct modeling under the binary source 
assumption: 1L2S model.

\begin{figure}
\includegraphics[width=\columnwidth]{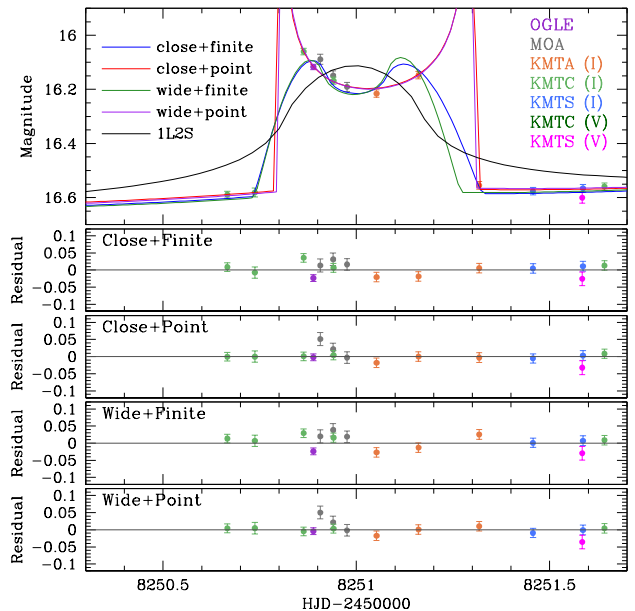}
\caption{
Comparison of model light curves in the region of the anomaly.  In the top 
panel, the curves marked in blue, red, green, and purple colors are models of 
the ``close+finite'', ``close+point'', ``wide+finite'', and ``wide+point'', 
respectively, which are the four degenerate solutions found from 2L1S modeling.  
The lower four panels show the residuals from the individual solutions.  The 
black curve is the solution obtained from the 1L2S modeling.  
\smallskip
}
\label{fig:two}
\end{figure}

\subsection{2L1S Modeling}

A short-term planetary microlensing signal is produced by the passage over 
or approach of the source close to the caustic induced by the planet. The 
planet-induced caustics are classified into two types: ``central'' and 
``planetary'' caustics. The central caustic is located close to the primary 
lens, while the planetary caustic is located away from the primary with a 
separation $\sim s-s^{-1}$.  For the properties of the planet-induced caustic, 
see  \citet{Chung2005} and \cite{Han2006} for the central and planetary 
caustics, respectively.

Under the planetary interpretation of the anomaly, one can heuristically characterize 
the planet.  The values of $(t_0, u_0, t_{\rm E}) \sim (8254.6, 0.035, 71~{\rm days})$ 
obtained from the 1L1S modeling
for the data excluding the anomaly, together with $t_{\rm anom}\sim 8251.0$, where
the times $t_0$ and $t_{\rm anom}$ are expressed in ${\rm HJD}^\prime\equiv {\rm HJD}-2450000$,
indicate that the caustic is located relatively close to the primary, and thus 
the perturbation is likely to be produced by the central caustic rather than 
the planetary caustic.  One can estimate the source trajectory angle $\alpha$ 
(with respect to the binary axis) from the relation
\begin{equation}
\alpha = \tan^{-1}\left( {u_0 t_{\rm E}\over t_{\rm anom}-t_0} \right)
\sim 2.5~{\rm radian},
\label{eq1}
\end{equation}
which is very similar to the value obtained from detailed modeling described 
below.

\begin{figure}
\includegraphics[width=\columnwidth]{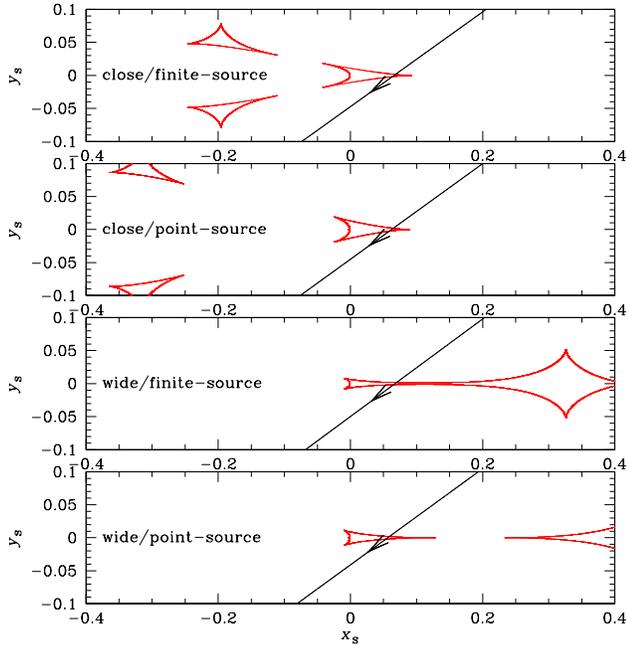}
\caption{
Lens-system configurations of the four degenerate solutions found from 2L1S 
modeling.  In each panel, the line with an arrow represents the source trajectory 
and the closed figures composed of concave curves are the caustics. 
The lensing parameters corresponding to the individual solutions are 
presented in Table~\ref{table:one}.
}
\label{fig:three}
\end{figure}

\begin{deluxetable*}{lcccc}
\tablecaption{Lensing Parameters of Planetary Solutions\label{table:one}}
\tablewidth{480pt}
\tablehead{
\multicolumn{1}{c}{Parameter}       &
\multicolumn{1}{c}{Close+Finite}    &
\multicolumn{1}{c}{Close+Point}     &
\multicolumn{1}{c}{Wide+Finite}     &
\multicolumn{1}{c}{Wide+Point}       
}
\startdata
$\chi^2$                    &   1265.4               &  1245.5               &  1267.5                 &  1244.4               \\
$t_0$ (${\rm HJD}^\prime$)  &   8254.347 $\pm$ 0.031 &  8254.242 $\pm$ 0.032 &  8254.351 $\pm$ 0.029   &  8254.223 $\pm$ 0.033 \\
$u_0$                       &   0.038    $\pm$ 0.003 &  0.036    $\pm$ 0.003 &  0.038    $\pm$ 0.003   &  0.033    $\pm$ 0.003 \\
$t_{\rm E}$ (days)          &   62.63    $\pm$ 4.37  &  64.58    $\pm$ 4.24  &  65.64    $\pm$ 3.47    &  70.12    $\pm$ 5.16  \\
$s$                         &   0.91     $\pm$ 0.01  &  0.86     $\pm$ 0.01  &  1.18     $\pm$ 0.01    &  1.26     $\pm$ 0.01  \\
$q$ ($10^{-3}$)             &   2.30     $\pm$ 0.18  &  4.28     $\pm$ 0.41  &  2.13     $\pm$ 0.17    &  4.54     $\pm$ 0.51  \\
$\alpha$ (rad)              &   2.520    $\pm$ 0.009 &  2.517    $\pm$ 0.009 &  2.508    $\pm$ 0.008   &  2.512    $\pm$ 0.010 \\
$\rho$ ($10^{-3}$)          &   0.99     $\pm$ 0.10  &  $< 0.4$              &  0.85     $\pm$ 0.08    &  $< 0.5$ 
\enddata
\tablecomments{${\rm HJD}^\prime\equiv {\rm HJD}-2450000$.
\bigskip
}
\end{deluxetable*}

We conduct systematic analysis of the observed data to search for the exact 
lensing parameters.  
In the first step 
of this analysis, we conduct grid searches for the 
binary-lens parameters $(s, q)$, while the other lensing parameters 
$(t_0, u_0, t_{\rm E}, \alpha)$ are searched for using a downhill method of 
the Markov Chain Monte Carlo (MCMC) algorithm \citep{Goodman2010}.
For the parameters $(t_0, u_0, t_{\rm E})$, we use the values obtained 
from the 1L1S modeling as initial parameters. 
For the source trajectory angle, we seed 
21 different initial values around a unit circle.  The sudden change of the 
source brightness ($\Delta I\sim 0.4$) before and after the perturbation 
suggests that the perturbation was produced by the caustic crossings of the 
source.  We, therefore, include an additional parameter of $\rho$ (normalized 
source radius), which represents the ratio of the angular source radius 
$\theta_*$ to $\thetae$, i.e., $\rho=\theta_*/\thetae$, to account for 
finite-source effects that affect the light curve during caustic crossings.  
From this first step analysis, we identify local minima in the $\Delta\chi^2$ plot on 
the plane of the grid parameters, i.e., $s$--$q$ plane.  In the second step, 
we refine the individual local solutions by allowing all parameters, both the 
grid parameters $(s,q)$ and the MCMC parameters $(t_0,u_0,t_{\rm E},\alpha, \rho)$, 
to vary.

From the 2L1S modeling, we identify four discrete degenerate solutions.
In Table~\ref{table:one}, we list the lensing parameters of these solutions,
together with their $\chi^2$ values.  The mass ratios for all of the solutions 
are $q< 10^{-2}$, indicating that the lens is a planetary system.  The planetary 
solutions greatly improves the fit by $\Delta\chi^2\sim 4630$ with respect to 
the 1L1S solution.  Despite the very strong planetary signal, the $\chi^2$ 
differences between the degenerate solutions are merely $\Delta\chi^2 \lesssim 23$, 
and thus the degeneracy is substantial.  In Figure~\ref{fig:two}, we present the 
model light curves of the individual solutions in the region of the anomaly.  
In Figure~\ref{fig:three}, we also present the lens-system configurations, 
which show the source trajectories with respect to the caustic, of the individual 
solutions.  As expected from the location of the anomaly lying close to the peak 
of the light curve and the strong deviation from the 1L1S model, the anomaly is 
produced by the crossing of the source over the central caustic induced by the 
planetary companion for all cases of the degenerate solutions.

\begin{figure}
\includegraphics[width=\columnwidth]{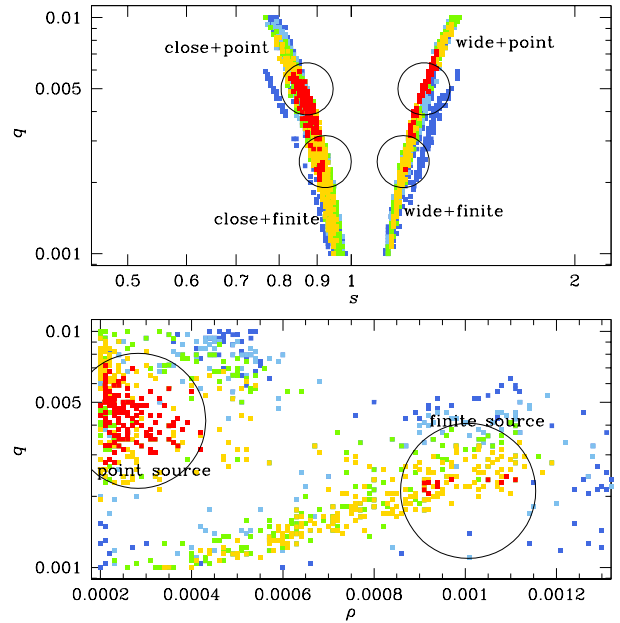}
\caption{$\Delta\chi^2$ distributions of points in the MCMC chain on the 
$s$--$q$ (upper panel) and $\rho$--$q$ (lower panel) planes. 
Red, yellow, green, cyan, and blue colors are used to indicate 
points with $1n\sigma$, $2n\sigma$, $3n\sigma$, $4n\sigma$, and $5n\sigma$, 
respectively, and $n=5$.
\bigskip
}
\label{fig:four}
\end{figure}

We inspect the origin of the degeneracy and find that there exist two types of 
degeneracy.  The first degeneracy is caused by the uncertain planet-host separation.  
This is shown in the upper panel of Figure~\ref{fig:four}, in which we present the 
$\Delta\chi^2$ distribution of MCMC points on the $s$--$q$ plane.  From 
the locations of the local solutions, it is found that two solutions have similar 
mass ratios of 
$q\sim 5\times 10^{-3}$ 
and the other two solutions have mass ratios 
$q\sim 2\times 10^{-3}$.  
For the pair of solutions with similar mass ratios, it is found that one solution 
has a separation $s<1.0$, while the other solution has a separation 
$s>1.0$.  This indicates that the degeneracy between the pair of solutions with 
$s<1.0$ and $s>1.0$ is caused by the well-known ``close/wide degeneracy''.  For 
planetary lens systems with very low mass ratios and projected separations substantially 
greater or smaller than unity, the planetary and central caustics are well separated.  
In such cases, the projected separations of the two degenerate solutions subject to the 
close/wide degeneracy are in the relation of $s\leftrightarrow s^{-1}$.  In the case of 
OGLE-2018-BLG-0740, the projected separations of the pairs of the degenerate solutions 
slightly deviate from this relation because the separations are close to unity: 
$s\sim 0.9$ for the close solutions and $s\sim 1.2$ for the wide solutions.

\begin{figure}
\includegraphics[width=\columnwidth]{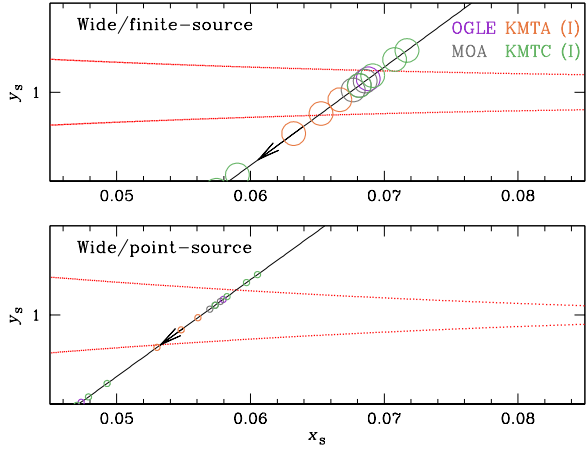}
\caption{
Source positions at around the time of the anomaly.  The upper and lower panels 
correspond to the ``close/finite-source'' and ``close/point-source'' solutions, 
respectively. In each panel, the line with an arrow is the source trajectory 
and the red curves represent the caustic.  
The small circles on the source trajectory represent the source size.  
Although the source is consistent with a point for the point-source solution, 
we show it with normalized radius $\rho=2.3\times 10^{-4}$ for reasons discussed 
in section~\ref{sec:six}.
\bigskip
}
\label{fig:five}
\end{figure}

We find that the other degeneracy arises due to the ambiguity in the normalized 
source radius, $\rho$, caused by the incomplete coverage of the anomaly.  To show 
this, we mark the positions of the local solutions on the $\rho$--$q$ parameter 
plane presented in the lower panel of Figure~\ref{fig:four}.  It is found that there 
exist two locals, in which the one with a smaller mass ratio, i.e., 
$q\sim 2\times 10^{-3}$, has a normalized source radius of 
$\rho\sim 1.0\times 10^{-3}$, while the normalized source radius of the other 
local with a larger mass ratio, i.e., $q\sim 5\times 10^{-3}$, is consistent with 
zero, i.e., point source. 
We refer to this degeneracy as the ``finite/point-source degeneracy''.  
The fact that the solutions with different $\rho$ values have different values 
of $s$ and $q$ indicates that the finite/point-source degeneracy causes ambiguity 
in the determinations of both $s$ and $q$, while the close/wide degeneracy causes 
ambiguity in the determination of only $s$.

\begin{figure}
\includegraphics[width=\columnwidth]{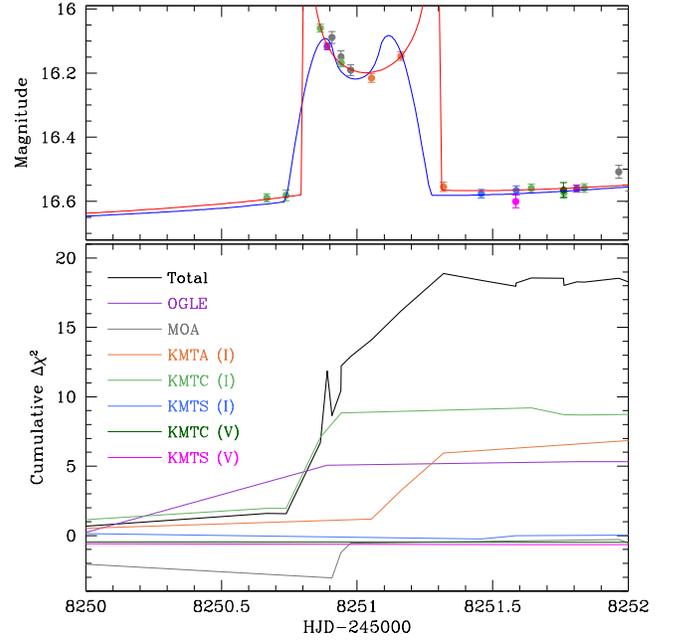}
\caption{
Cumulative distribution of $\Delta\chi^2$ 
between the ``finite'' and ``point'' solutions (with $s<1.0$) 
in the region of the anomaly.
The black curve is the distribution for all data sets, and 
the distributions for the individual data sets are marked in
different colors that match those of the legends.
The light curve in upper panel is presented to show the region of 
$\chi^2$ difference.
\bigskip
}
\label{fig:six}
\end{figure}

We designate the individual local 
solutions as ``close+finite'', ``close+point'', ``wide+finite'', and ``wide+point''. 
Here the terms ``close'' and ``wide'' indicate $s<1$ and $s>1$, respectively.  The 
term ``finite'' is used to represent that the light curve is subject to finite-source 
effects, while the term ``point'' is used to represent that the light curve is 
consistent with that of a point-source event.  According to the point-source solutions, 
the source positions corresponding to the anomalous data points are well within the 
caustic and the data points during the anomaly are placed in the ``U''-shape trough 
region between the caustic-crossing spikes of the light curve.  According to the 
finite-source solutions, on the other hand, most data points correspond to the 
source positions during which the source was crossing the caustic.  See the data 
points around the anomaly region of the light curve presented in 
Figure~\ref{fig:two} and the corresponding source positions presented in 
Figure~\ref{fig:five}.

\begin{figure}
\includegraphics[width=\columnwidth]{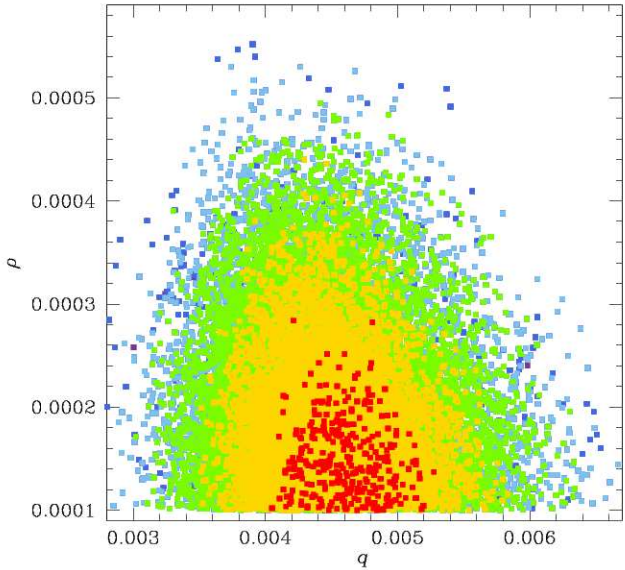}
\caption{
$\Delta\chi^2$ distribution of points in the MCMC chain on the $q-\rho$ plane.
for the 
``wide+point'' 
solution. 
Color coding is set to represent points within $1\sigma$ (red), $2\sigma$ (yellow), 
$3\sigma$ (green), $4\sigma$ (cyan), and $5\sigma$ (blue) from the best-fit value.  
\bigskip
}
\label{fig:seven}
\end{figure}

In order to see the severity of the ``finite/point-source'' degeneracy, in 
Figure~\ref{fig:six}, we present the cumulative distribution of $\chi^2$ 
difference between the ``finite-source'' and ``point-source'' solutions 
in the region of the anomaly.   
We note that the plot is for the pair of the wide solutions with $s>1.0$, 
but for the pair of the close solutions with $s<1.0$,
the plot is very similar to the presented $\Delta\chi^2$ distribution.
From the $\Delta\chi^2$ plot, it
is found that the point-source solution is favored over the finite-source solution 
by $\Delta\chi^2\sim 18$ during the anomaly.  However, considering that this level 
of $\Delta\chi^2$ might be caused by systematics in microlensing data together with 
the fact that the $\chi^2$ difference originates mainly from a few points, it is 
difficult to completely rule out the finite-source solution just based on the 
observed photometric data.

We inspect the higher-order effects in the observed light curve caused by the 
orbital motion of the observer, microlens-parallax effects \citep{Gould1992a}, 
and the orbital motion of the lens, lens-orbital effects \citep{Dominik1998}.  
For this inspection, we conduct a series of modeling separately and simultaneously 
considering these effects.  We find that these modeling runs result in little 
improvement of the fit, with $\Delta\chi^2\sim 2$ when both the higher-order effects 
are simultaneously considered.  The difficulty of measuring the higher-order effects 
arises because the source is a faint star with $I\sim 21.5$ (see section~\ref{sec:four}), 
and thus the photometric quality is not high enough to detect subtle deviations induced 
by the higher-order effects, despite the relatively long timescale, 
$t_{\rm E}\gtrsim 60$ days, of the event.

\subsection{1L2S Modeling}

Because it is known that a short-term anomaly can also be produced 
by a binary companion to the source, we additionally conduct a 1L2S modeling.  
Besides the 1L1S lensing parameters of 
$(t_0, u_0, t_{\rm E})$, this modeling requires to include additional 
parameters of $(t_{0,2}, u_{0,2}, q_F)$, where $t_{0,2}$ is the time of the 
closest lens approach to the source companion, $u_{0,2}$ is the lens-companion 
separation at $t_{0,2}$, and $q_F$ represents the flux ratio between the two 
source stars.  For the initial value of $t_{0,2}$, we use the time of the 
anomaly, $t_{\rm anom}$.  We set the initial values of $u_{0,2}$ and $q_F$ 
considering that the source has a small flux ratio, $q_F \ll 1$, and the lens 
approaches very close to the source companion, $u_{0,2}\rightarrow 0$, for 
1L2S events producing short-term anomalies.

In the upper panel of Figure~\ref{fig:two}, we present the model light curve 
of the best-fit 1L2S solution.  It is found that the model provides a poorer 
fit than the 2L1S solutions.  The $\chi^2$ difference between the 1L2S and 
2L1S solutions is $\Delta\chi^2\sim 606$.  We, therefore, reject this 
interpretation of the anomaly and conclude that the origin of the anomaly is 
the planetary companion to the lens.

\section{Source and Blend}\label{sec:four}

We characterize the source star based on its de-reddened color $(V-I)_0$ and 
brightness $I_0$.  For the determinations of $(V-I)_0$ and $I_0$, we use 
the method of \citet{Yoo2004}, which utilizes the centroid of the red giant clump 
(RGC) in the color-magnitude diagram (CMD) as a reference to calibrate the color 
and brightness of the source star.  Defining the source star is important for 
the determination of the angular Einstein radius because $\thetae$ is related 
to the angular radius of the source star, $\theta_*$, by the relation
\begin{equation}
\thetae={\theta_*\over \rho}, 
\label{eq2}
\end{equation}
where $\theta_*$ is estimated from the source type and the normalized source 
radius $\rho$ is measured by analyzing the 
caustic-crossing parts of the light curve.

\begin{figure}
\includegraphics[width=\columnwidth]{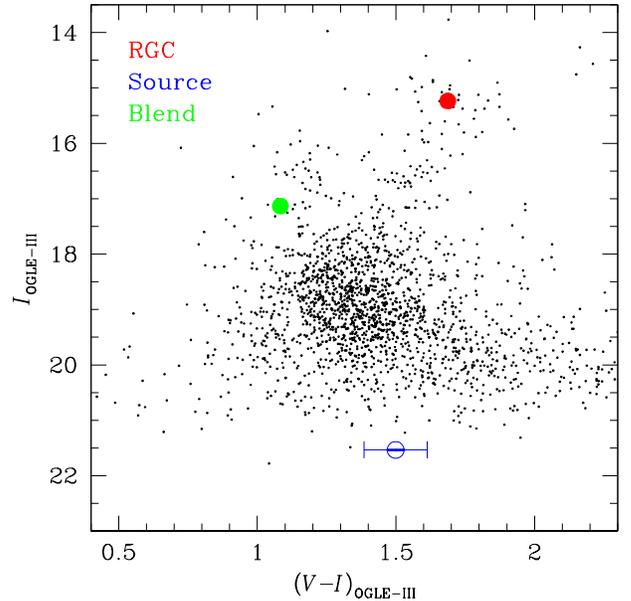}
\caption{
Locations of the source and blend in the color-magnitude diagram with respect 
to the centroid of red giant clump (RGC).  The color and magnitude are estimated 
based on the KMTC data set, but they are calibrated to OGLE-III photometry.
\bigskip
}
\label{fig:eight}
\end{figure}

\begin{figure*}
\epsscale{0.90}
\plotone{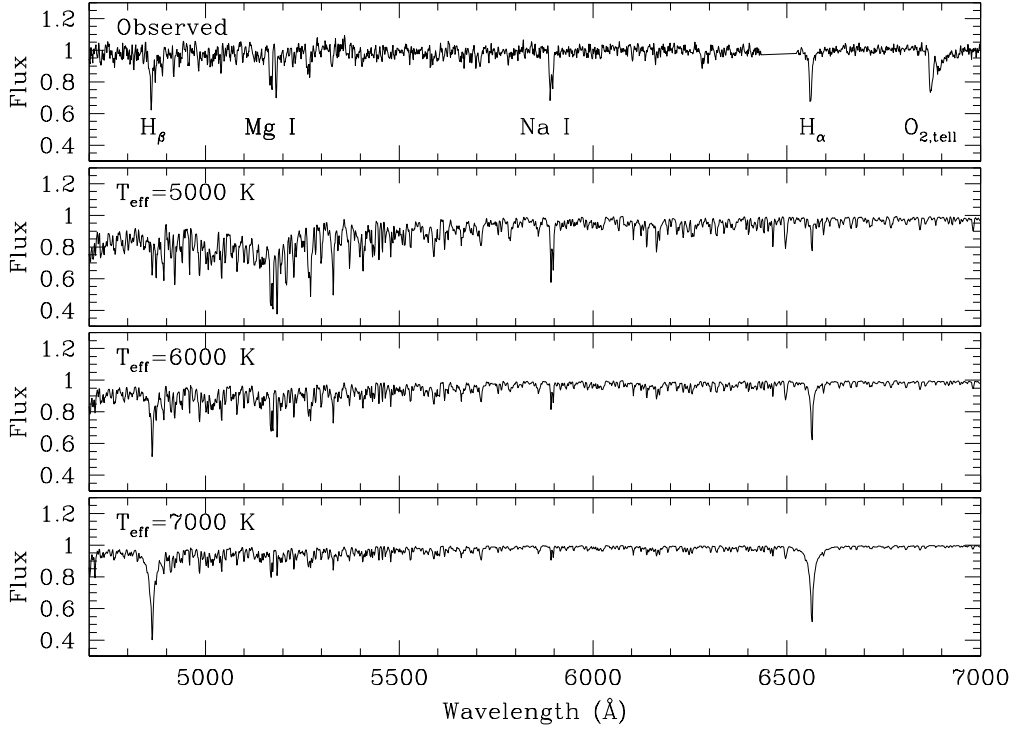}
\caption{
Observed spectrum of the baseline object (top panel).
The three lower panels show the synthetic spectra of stars 
with $T_{\rm eff}=5000$~K, $T_{\rm eff}=6000$~K, and $T_{\rm eff}=7000$~K,
which correspond to those of early 
K, G, and F-type stars, respectively.
We mark the major spectral lines used to determine the spectral type.
The line marked by ``${\rm O}_{\rm 2,tell}$'' (telluric ${\rm O}_2$ line) indicates the 
oxygen molecular line produced by the Earth's atmosphere.
\bigskip
}
\label{fig:nine}
\end{figure*}

We note that the angular Einstein radius can be measured from the relation in 
Equation~(\ref{eq2}) for the ``finite-source'' solutions because the $\rho$ 
value is measured for these solutions, but the value of $\rho$ is not measured 
for the ``point-source'' solutions and thus $\thetae$ cannot be measured.  For 
the ``point-source'' solutions, however, one can set the lower limit of the 
angular Einstein radius by
\begin{equation}
\theta_{\rm E,min}={\theta_*\over \rho_{\rm max}}, 
\label{eq3}
\end{equation}
where $\rho_{\rm max}$ represents the upper limit of the normalized source radius.
In Figure~\ref{fig:seven}, we present the $\Delta\chi^2$ distribution of the points 
in the MCMC chain on the $q-\rho$ plane for the ``wide+point'' solution.  It shows 
that the upper limit is $\rho_{\rm max}\sim 0.5\times 10^{-3}$ as measured 
at the $3\sigma$ level.  The ``close+point'' solution yields a similar value of 
$\rho_{\rm max}$.

In Figure~\ref{fig:eight}, we mark the location of the source (empty circle 
with error bars) in the CMD of stars around the source.  Also marked are the 
locations of the RGC centroid (red dot) and blend (green dot).  We determine 
the $I$- and $V$-band magnitudes of the source using the regression method for 
the KMTC photometry data processed with the pyDIA photometry.  We note that 
the KMTC data are not calibrated, and thus the color and magnitude are scaled 
to those of OGLE-III CMD \citep{Szymanski2011} in order to estimate calibrated 
color and brightness.  The apparent color and brightness of the source are  
$(V-I,I) = (1.49\pm 0.11, 21.54\pm 0.01)$.  From the offset of 
$\Delta (V-I,I) = (-0.20, 6.30)$ with respect to the RGC centroid at 
$(V-I,I)_{\rm RGC} = (1.69, 15.24)$, together with the known de-reddened color 
and brightness of the RGC centroid of $(V-I,I)_{\rm RGC,0}=(1.06, 14.35)$ 
\citep{Bensby2011, Nataf2013}, we estimate that the de-reddened color and 
brightness of the source stars are $(V-I,I)_0 = (0.87\pm 0.11, 21.53\pm 0.01)$.  
This indicates that the source is an early K-type main-sequence star.  Once 
the de-reddened $V-I$ color is measured, we then convert $V-I$ color 
into $V-K$ color using the color-color relation of \citet{Bessell1988}, and 
then estimate the angular radius of the source using the $(V-K)/\theta_*$ 
relation of \citep{Kervella2004}.  The estimated angular source radius from 
this procedure is 
\begin{equation}
\theta_*=0.28 \pm 0.04~\mu{\rm as}.
\label{eq4}
\end{equation}

With the measured angular source radius, the angular Einstein radius, $\thetae$, 
and the relative lens-source proper motion, $\mu=\thetae/t_{\rm E}$, are 
determined.  These values are 
\begin{equation}
\theta_{\rm E, FS}=0.28 \pm 0.04~{\rm mas},\qquad
\mu_{\rm FS} = 1.65 \pm 0.23~{\rm mas}~{\rm yr}^{-1}
\label{eq5}
\end{equation}
for the ``finite-source'' solutions.  We note that the estimated values of $\thetae$ 
and $\mu$ for the finite-source solutions are substantially lower than 
$\langle \thetae \rangle\sim 0.5~{\rm mas}$ and 
$\langle \mu\rangle \sim 5~{\rm mas}~{\rm yr}^{-1}$ of typical Galactic lensing 
events produced by low-mass stars, $\sim 0.3~M_\odot$, located halfway between 
the source and observer, $D_{\rm L}\sim  4$~kpc, in the disk.  If the finite-source 
solution were correct, then the lens would very likely lie in the bulge both because 
of the small Einstein radius and low proper motion.

For the ``point-source'' solutions, for which only the upper limit of $\rho$ is 
determined, the corresponding lower limits are
\begin{equation}
\theta_{\rm E, PS} > 0.56~{\rm mas},\qquad
\mu_{\rm PS} > 2.9~{\rm mas}~{\rm yr}^{-1}.
\label{eq6}
\end{equation}
These values are reasonably consistent with those of typical disk lens events.  
See section~\ref{sec:five} for the detailed discussion of the probable lens 
locations for the individual solutions based on microlensing data alone, 
and see section~\ref{sec:six} for the final determination of the lens distance 
using external data.  In Table~\ref{table:two}, we list the estimated values 
of $\thetae$ and $\mu$ for the finite-source and point-source solutions.

The observed flux is dominated by blended light and thus we also characterize the 
blend.  Another reason for identifying the blend is to check the possibility that 
the blend is the lens such as in the case of  OGLE-2017-BLG-0039 \citep{Han2018b}.  
The apparent color and brightness of the blend for OGLE-2018-BLG-0740 
are $(V-I,I)_{\rm b}=(1.08, 17.13)$.  Considering the color and brightness, the 
blend is likely to be a main-sequence star located in the disk.  The extinction 
and reddening toward the field are $A_I\sim 0.79$ and $E(V-I)\sim 0.66$, respectively 
\citep{Gonzalez2012}.  Assuming that the blend experiences $\sim 1/2$ of the total 
extinction and reddening toward the bulge field, the de-reddened color of the blend 
is estimated as $(V-I)_{0,{\rm b}}\sim (V-I)_{\rm b} - E(V-I)/2 \sim 0.75$.  This 
corresponds to the color of a G-type star.

The dominance of the observed flux by the blended light combined with the bright 
nature of the blend object enable us to characterize the blend based on the spectra 
obtained from follow-up observations.  The spectroscopic follow-up observations were 
conducted using the IMACS spectrograph of the Magellan-Baade telescope on UT 2019-03-22 
08:40 with 5 min of exposure. We chose the 300 lines/nm grism with 17.5 degrees of Blaze 
Angle (i.e., dispersion of 1.341~\AA/pixel) of the f/2 channel and the 0.9 arcsec slit.  
In Figure~\ref{fig:nine}, we present the spectrum of the baseline object along with the 
synthetic spectra of stars from the BOSZ spectral library37 with $T_{\rm eff} = 5000$~K, 
$T_{\rm eff} = 6000$~K, and $T_{\rm eff} = 7000$~K, which correspond to those of early 
K, G, and F-type stars, respectively. By fitting the blue portion of the IMACS spectrum 
as a function of parameters interpolated over the ELODIE 3.2 library \citep{Moultaka2004} 
of stellar spectra using the University of Lyon Spectroscopic analysis Software (Ulyss) 
code \citep{Koleva2009, Wu2011}, we find that $T_{\rm eff} = 5912 \pm 49$~K, 
$\log g = 4.5 \pm 0.1$, and $[{\rm Fe}/{\rm H}] = -0.24 \pm 0.05$, indicating that the 
blend is an early G-type star, very similar to the Sun in temperature and gravity, 
though slightly lower in metallicity. This is consistent with the spectral type estimated 
based on the photometric data.

\section{Physical Lens Parameters (Bayesian analysis)}\label{sec:five}

The mass $M$ and distance $D_{\rm L}$ to a lens can be uniquely determined 
when both the angular Einstein radius $\thetae$ and the microlens parallax $\pie$ 
are simultaneously measured, i.e.,
\begin{equation}
M={\thetae\over \kappa\pie};\qquad
D_{\rm L}={{\rm au}\over \pie\thetae +\pi_{\rm S}},
\label{eq7}
\end{equation}
where $\kappa=4G/(c^2{\rm au})$, $\pi_{\rm S}={\rm au}/D_{\rm S}$, and $D_{\rm S}$ 
denotes the distance to the source.  For OGLE-2018-BLG-0740, the angular Einstein radius is 
measured for the finite-source solutions and the lower limit is constrained for 
the point-source solutions, but the microlens-parallax is not measured for 
either of these solutions.  We, therefore, estimate the physical lens parameters 
by conducting a Bayesian analysis of the event with the constraints of the measured 
event timescale and angular Einstein radius.

The Bayesian analysis is carried out by producing a large number of events, $10^{6}$,
from a Monte Carlo simulation based on the prior conditions of lens mass composition, 
i.e., mass function, and the distributions of astronomical objects and their motion, 
i.e., physical and dynamical distributions, respectively.  For the mass function, we 
adopt the \citet{Chabrier2003} model for stars and the \citet{Gould2000} model for 
stellar remnants.  We adopt the \citet{Han2003} model for the physical distribution 
of matter in the Galaxy and the \citet{Han1995} model for the dynamical distribution.  
More detailed description of the adopted prior models is found in section 5 of 
\citet{Han2018a}.  From the probability distributions of the physical parameters 
for events with timescales and Einstein radii within the ranges of the measured 
values, we estimate the physical parameters and their uncertainties.

\begin{figure}
\includegraphics[width=\columnwidth]{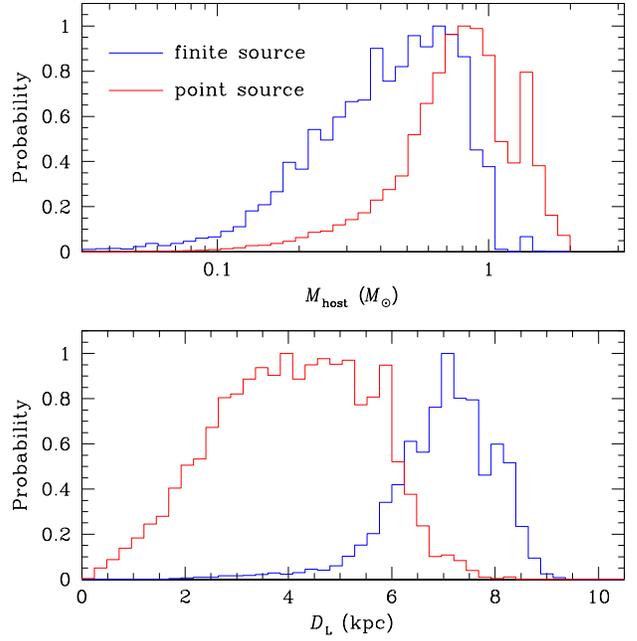}
\caption{
Probability distributions of the lens mass
(upper panel) and distance (lower panel). The blue
and red curves are the distributions for the finite
and point-source solutions, respectively.
\smallskip
}
\label{fig:ten}
\end{figure}

In Figure~\ref{fig:ten}, we present the probability distributions of the host 
mass and the distance to the lens constructed from the Bayesian analysis.  Because 
there exist two classes of degenerate solutions, i.e., finite-source and point-source 
solutions, we present two probability distributions corresponding to the individual 
classes of solutions.  In Table~\ref{table:two}, we present the estimated physical 
parameters of the lens system, including the host mass, $M_{\rm host}$, planet mass, 
$M_{\rm p}$, distance to the lens, $D_{\rm L}$, and the physical projected separation 
between the host and planet, $a_\perp$.   We note that the presented physical parameters 
correspond to the median values of the probability distributions  and their lower and 
upper limits are estimated as the 16\% and 84\% of the distributions, respectively.  
We note that the two values of $a_\perp$ correspond to the close and wide solutions.

To be noted about the lens parameters is that the physical parameters 
estimated from the two classes of the degenerate solutions are substantially 
different.  
According to the point-source solutions, the masses of the host 
and planet are
\begin{equation}
M_{\rm host}=0.83^{+0.47}_{-0.33}~M_\odot,
\label{eq8}
\end{equation}
and 
\begin{equation}
M_{\rm p}=3.9^{+2.2}_{-1.6}~M_{\rm J},
\label{eq9}
\end{equation}
respectively, and the distance to the lens is 
\begin{equation}
D_{\rm L}= 4.2^{+1.6}_{-1.6}~{\rm kpc}.
\label{eq10}
\end{equation}
In this case, the planetary system consists of a super-Jupiter mass planet and 
a G--K type host star located more likely in the disk.  
According to the finite-source solutions, on the other hand, 
the masses of the 
lens components are  
\begin{equation}
M_{\rm host}= 0.47^{+0.31}_{-0.24}~M_\odot
\label{eq11}
\end{equation}
and 
\begin{equation}
M_{\rm p}=1.1^{+0.7}_{-0.6}~M_{\rm J},
\label{eq12}
\end{equation}
respectively. The lens is located at a distance of 
\begin{equation}
D_{\rm L}=7.2^{+0.9}_{-1.0}~{\rm kpc}.
\label{eq13}
\end{equation}
Then, the lens is a planetary system composed of a planet with a mass similar 
to that of Jupiter and an M-dwarf host located in the Galactic bulge.
The differences between 
the physical parameters for the two classes of solutions indicate that the 
degeneracy reported in this work greatly affects the determinations of the 
lens parameters.

\begin{deluxetable}{llcc}
\tablecaption{Lens Parameters from Bayesian Analysis\label{table:two}}
\tablewidth{240pt}
\tablehead{
\multicolumn{1}{c}{Parameter}      &
\multicolumn{1}{c}{Point-source}   &
\multicolumn{1}{c}{Finite-source}       
}
\startdata
$\thetae$ (mas)               &  $> 0.56$                        &  0.28 $\pm$ 0.04                 \\
$\mu$ (mas~yr$^{-1}$)         &  $> 2.9$                         &  1.65 $\pm$ 0.23                 \\
$M_{\rm host}$ ($M_\odot$)    &  $0.83^{+0.47}_{-0.33}$          &  $0.47^{+0.31}_{-0.24}$          \\
$M_{\rm p}$ ($M_{\rm J}$)     &  $3.9^{+2.2}_{-1.6}$             &  $1.1^{+0.7}_{-0.6}$             \\
$D_{\rm L}$ (kpc)             &  $4.2^{+1.6}_{-1.6}$             &  $7.2^{+0.9}_{-1.0}$             \\
$a_\perp$ (au)                &  $3.3^{+1.3}_{-1.3}$ (close)     &  $1.5^{+0.2}_{-0.2}$ (close)     \\
                              &  $4.6^{+1.8}_{-1.8}$ (wide)      &  $2.0^{+0.2}_{-0.3}$ (wide)      
\enddata
\tablecomments{ 
The two values of the projected planet-host separation $a_\perp$ correspond 
to the close and wide solutions.
\smallskip
}
\end{deluxetable}

\section{Resolving the Degeneracy}\label{sec:six}

Below we show that the degeneracy between the point-source and finite-source solutions 
is resolved in favor of the point-lens solution.  In addition to the slightly 
better fit to the data, it is found that the point-source solution is also
supported by the external information obtained from astrometric and spectroscopic 
observations.

One line of evidence for the preference of the point-source solution comes from 
the identification of the blend with either the lens or its companion.  This is 
identified from the astrometric measurement of the offset between the source and 
baseline object in the KMTNet images.  From this measurement, it is found that the 
offset is 0.07 pixels, which corresponds to $\sim 28$~mas.  The uncertainty in the 
position of the source measured on the difference image is $\sim 0.1$ pixel, and 
thus the measured astrometric offset is consistent with the measurement uncertainty 
of the baseline object on the template.  The probability of a random field star with 
a brightness similar to or brighter than that of the blend with $V_{\rm b}=18.21$ 
lying this close to the source is extremely low, $p=3.6\times 10^{-5}$.  This indicates 
that the blend is not a field star that happens to lie close to the source.  Excluding 
the possibility of a random field star, then the blend can only be the lens itself, 
a companion to the lens, or a companion to the source.  However, the blend cannot be 
a companion to the source because its spectroscopic type and observed color and 
magnitude place it well in the foreground.  More specifically, being just slightly 
hotter than the Sun, with marginally higher surface gravity, the blend has 
$(V-I)_0\sim 0.68$ and $M_I\sim 4.1$.  Hence, 
$E(V-I)=(V-I)_{\rm b}-(V-I)_{\rm 0,b}=1.08-0.68=0.4$ and so $A_I\simeq 1.25 E(V-I)=0.5$.  
These values imply a distance modulus ${\rm DM}=I_{\rm L} - M_{I,{\rm L}} - A_{I,{\rm L}} 
\simeq 12.5$, and thus the blend is located at a distance
\begin{equation}
D_{\rm b} = 10^{({\rm DM}+5)/5-3} 
= 3.2\pm 0.5~{\rm kpc}. 
\label{eq14}
\end{equation}

\begin{deluxetable}{lcc}
\tablecaption{Lens Parameters from Spectrum\label{table:three}}
\tablewidth{240pt}
\tablehead{
\multicolumn{1}{c}{Parameter}       &
\multicolumn{1}{c}{close}           &
\multicolumn{1}{c}{wide}       
}
\startdata
$\thetae$ (mas)               &  $1.24 \pm 0.15$     &   --                      \\
$\mu$ (mas yr$^{-1}$)         &  $7.0 \pm 0.9$       &   $6.5 \pm 0.8$           \\
$M_{\rm host}$ ($M_\odot$)    &  $1.0 \pm 0.1$       &   --                      \\
$M_{\rm p}$ ($M_{\rm J}$)     &  $4.5 \pm 0.6$       &   $4.8 \pm 0.6$           \\
$D_{\rm L}$ (kpc)             &  $3.2 \pm 0.5$       &               \\
$a_\perp$ (au)                &  $3.4 \pm 0.2$       &   $5.0 \pm 0.3$           \\
$a$ (au)                      &  $4.2 \pm 0.6$       &   $6.1 \pm 0.6$           \\
$P$ (yr)                      &  $8.6 \pm 1.9$       &   $15.0 \pm 3.2$          \\ 
$v\sin i$ (${\rm m~s}^{-1}$)  &  $(62 \pm 11)\sin i$ &   $(55 \pm 10)\sin i$     \\
$I_{\rm L}$ (mag)             & $17.13 \pm 0.01$     &  --                      \\
$V_{\rm L}$ (mag)             & $18.21 \pm 0.11$     &  --   
\enddata
\tablecomments{ 
$\thetae$ and $\mu$ denote the angular Einstein radius and the relative lens-source
motion, respectively.  $a_\perp$ and $a$ represent the projected planet-host separation 
at the time of the peak lensing magnification and the semi-major axis, respectively.  
$P$ denotes the orbital period and $v\sin i$ represents the RV amplitude.  $I_{\rm L}$ 
and $V_{\rm L}$ represent the $I$ and $V$-band magnitudes of the lens, respectively.  
\bigskip
}
\end{deluxetable}

We now argue that the blend is very likely to be the lens itself rather than 
its companion.  The first point is that Equation~(\ref{eq14}) implies that 
the lens-source relative parallax is
\begin{equation}
\pi_{\rm rel}={\rm au}~\left({1\over D_{\rm L}}-{1\over D_{\rm S}}\right) 
= 0.188\pm 0.042~{\rm mas}.
\label{eq15}
\end{equation}
Here we adopt the distance to the source star of $D_{\rm S}=8$~kpc.
This is also the ``blend-source relative parallax''. Hence we can define an 
``Einstein radius of the blend''
\begin{equation}
\theta_{\rm E,b} = 
(\kappa M_{\rm b} \pi_{\rm rel})^{1/2} 
= 1.24\pm 0.15~{\rm mas},
\label{eq16}
\end{equation}
where $M_{\rm b}$ denotes the mass of the blend.
If the lens is the blend, then $\theta_{\rm E,b}=\thetae$.  But if not, then 
$\theta_{\rm E,b}$ is still a useful concept.
Now, suppose that the blend is a companion to the lens. We know that the mass 
ratio $Q\equiv M_{\rm b}/M_1\gg 1$ because the spectrum does not show significant 
light from a second star.  Suppose that the lens and blend are separated by 
$\Delta\theta$. We define $S\equiv \Delta\theta/\thetae$, where again, $\thetae$ 
is the Einstein radius of the lens. Then the semi-diameter of the Chang-Refsdal 
(CR) caustic (in units of $\thetae$) is $w=2Q/S^2$. We know that strictly 
$w < u_0=0.04$, because otherwise the light curve in Figure~\ref{fig:one} would 
show huge residuals near the peak caused by the CR caustic. We estimate 
$w < u_0/2$ to avoid detectable residuals. This leads to a limit 
\begin{equation}
\Delta\theta_{\rm min}=
S_{\rm min}\thetae
=\sqrt{{4Q \theta_{\rm E}^{2}\over u_0}}
=\sqrt{{4\over u_0}} \theta_{\rm E,b} \sim 12.4~{\rm mas},
\label{eq17}
\end{equation}
because $\theta_{\rm E,b}=\sqrt{Q}\theta_{\rm E}$.  We note that 
$\Delta\theta_{\rm max}$ is independent of the lens mass $M_1$.  But we also 
know that $\Delta\theta \lesssim 28$~mas, which is the measured astrometric 
offset between the source and baseline object.  Hence, the allowed range in 
separation is about 1/2 dex, which corresponds to 3/4 dex in period, which is 
centered on roughly $\log (P/{\rm day})\sim 5.2$.  From \citet{Duquennoy1991}, 
about 6\% of G dwarfs have companions in this separation range. Again using 
\citet{Duquennoy1991} statistics, the mean Einstein radius (hence cross section) 
of these companions will be lower than the primary by a factor 0.64. Hence, there 
is only a $0.64\times6\%\simeq 4\%$ chance that the lens is a companion to the 
blend rather than the blend itself.
Finally, we consider the specific case of the ``finite source'' solutions, for 
which $\thetae=0.28$~mas (Eq.~\ref{eq5}).  Because 
$\thetae \ll \theta_{\rm E,b} =\sqrt{Q}\thetae$, the lensing object responsible 
for the event is the lower-mass component of the binary with 
$Q=(\theta_{\rm E,b}/\thetae)^2=(1.24~{\rm mas}/0.28~{\rm mas})^2\sim 20$.  As 
we just argued, the general probability that the lens is such a companion is 
low ($\sim 5\%$). In addition, this would be a very unusual 3-body system, i.e., 
a solar-mass star, a brown-dwarf companion at 50--100~au, orbited by a 
two-Neptune-mass ``moon'' at about 1~au. Given that this model is already 
seriously disfavored by the microlensing data, we regard the low statistical 
probability just reported as well as the extreme nature of the system implied 
as essentially ruling out this possibility.  Then, the only remaining possibility 
is that the blend is the lens itself.

Knowing that the blend is very likely to be the lens, another line of 
evidence supporting the point-source solution comes from the consistency 
of the external distance measurement of the lens, i.e., blend, by spectrum with 
the lensing parameters of the point-source solution.  With the angular 
Einstein radius of $\thetae\simeq 1.24$~mas (Eq.~\ref{eq16}) together 
with the estimated source radius of $\theta_*\simeq 0.28~\mu{\rm as}$
(Eq.~\ref{eq4}), the normalized source radius is 
\begin{equation}
\rho={\theta_*\over \thetae}\sim 2.3 \times 10^{-4}.
\label{eq18}
\end{equation}
The estimated value of $\rho$ is consistent with the centroid of the 
cloud of MCMC points in the $\Delta\chi^2$ distribution on the $q$--$\rho$ 
plane for the point-source solution presented in Figure~\ref{fig:seven}.  
This indicates that the spectroscopically estimated lens distance is consistent 
with the lensing parameters of the point-source solution.  
In contrast, the spectroscopically estimated value of $\rho$ (Eq.~\ref{eq18})
is significantly different from the value of the finite-source solutions, 
which is located in the range $8.5\times 10^{-4} \lesssim \rho\lesssim 9.9\times 10^{-4}$.
We note that OGLE-2018-BLG-0740 is the first external mass measurement of a 
microlens by spectrum, and this result is consistent with all microlens model 
information.

Although consistent, the physical lens parameters estimated from the 
spectrum are slightly different from those estimated from the Bayesian 
analysis.  Therefore, we additionally list the lens parameters based on 
the spectrum in Table~\ref{table:three}.

\begin{deluxetable}{lc}
\tablecaption{Annual Parallax and Proper Motion \label{table:four}}
\tablewidth{240pt}
\tablehead{
\multicolumn{1}{c}{Parameter}       &
\multicolumn{1}{c}{Value}       
}
\startdata
$\pi$ (mas)                  & $0.27\pm 0.19$       \\
$\mu_{E}$ (mas~yr$^{-1}$)    & $1.67\pm 0.36$       \\
$\mu_{N}$ (mas~yr$^{-1}$)    & $-4.85\pm 0.31$    
\enddata
\tablecomments{ 
The values are adopted from the {\it Gaia} archive.
The quantity $\pi$ denotes the annual parallax, and $\mu_{E}$ and $\mu_{N}$ 
represent the east and north components of the proper motion, respectively.
\bigskip
}
\end{deluxetable}

\section{Radial-velocity Follow-Up Observation}\label{sec:seven}

The facts that 
(1) the blend is the lens,
(2) the lens, with $I_{\rm L}\simeq 17.1$ and $V_{\rm L}\simeq 18.2$,
is substantially brighter than typical lenses, and
(3) its flux dominates the observed flux
suggest that extra information such as the period and eccentricity of the 
planetary system can be additionally obtained from follow-up RV observations.  
In this section, we estimate the expected RV amplitude of the planetary system 
for future follow-up observations.

With the spectroscopically determined mass of the planet host, i.e., $M\sim 1.0~M_\odot$, 
together with the planet/host mass ratios of the point-source solutions, 
the mass of the planet is 
\begin{equation}
M_{\rm p}= q M_{\rm host} =
\begin{cases}
 4.5\pm 0.6~M_{\rm J} & {\rm (close)}, \\
 4.8\pm 0.6~M_{\rm J} & {\rm (wide)},
\end{cases}
\label{eq19}
\end{equation}
where the upper and lower cases represent the values corresponding to the 
close and wide solutions, respectively.  The projected host-planet separation is 
\begin{equation}
a_\perp = s\thetae D_{\rm L} =
\begin{cases}
 3.4\pm 0.2~{\rm au} & {\rm (close)}, \\
 5.0\pm 0.3~{\rm au} & {\rm (wide)}.
\end{cases}
\label{eq20}
\end{equation}
Assuming a circular orbit and 
a random orientation of the planet around the host, the mean value of the
intrinsic semi-major axis is
\begin{equation}
\langle a \rangle = \sqrt{{3\over 2}} a_\perp =
\begin{cases}
 4.2\pm 0.6~{\rm au} & {\rm (close)}, \\
 6.1\pm 0.6~{\rm au} & {\rm (wide)}.
\end{cases}
\label{eq21}
\end{equation}
From the Kepler's third law, the orbital period of the planet is 
\begin{equation}
P = \left({a^3\over M}\right)^{1/2} \sim 
\begin{cases}
 8.6 \pm 1.9~{\rm yr} & {\rm (close)}, \\
 15.0\pm 3.2~{\rm yr} & {\rm (wide)}.
\end{cases}
\label{eq22}
\end{equation}
Then, the expected RV amplitude is
\begin{equation}
v\sin i = q\left({2\pi a\over P}\right)\sin i  \sim 
\begin{cases}
 (62\pm 11)\sin i~{\rm m~s}^{-1} & {\rm (close)}, \\
 (55\pm 10)\sin i~{\rm m~s}^{-1} & {\rm (wide)}.
\end{cases}
\label{eq23}
\end{equation}
These RV amplitudes are big enough to be measured using high-resolution 
spectrometers mounted on very large telescopes.  For example, for a G-type 
star at $V=18$, VLT/Espresso can achieve $10~{\rm m~s}^{-1}$ precision with 
a single VLT telescope and $5$--$10~{\rm m~s}^{-1}$ precision by employing all 
the 4 VLT telescopes. 
To be noted is that the RV signal, i.e., $v \sin i$, depends on the 
inclination of the planet orbit. 
For detections, therefore, the planet would need to have a
large inclination, as pointed out 
by \citet{Clanton2014}.

The motion of the lens is defined from the combination of spectroscopic and 
astrometric data.  The radial velocity is measured from the Magellan spectrum as 
$v_r = -36\pm 5~{\rm km~s}^{-1}$ after heliocentric correction.  The projected 
velocity is estimated from the proper motion and distance by ${\bf v}=\muvec D_{\rm L}$.  
In Table~\ref{table:four}, we list the proper motion, $\muvec=(\mu_E, \mu_N)$, and 
annual parallax, $\pi$, of the lens 
from the list of {\it Gaia} data release 2 \citep[{\it Gaia} DR2:][]{Gaia2018}.
Then, the east and north components of the projected lens velocity are 
$(v_E,v_N)=(\mu_E, \mu_N) D_{\rm L}= (25\pm 5, -73\pm 5)~{\rm km~s}^{-1}$.  
We note that the distance to the lens, 
$2.2~{\rm kpc} \leq  D_{\rm L}\equiv 1/\pi \leq 12.5~{\rm kpc}$,
estimated from the {\it Gaia} annual parallax, i.e., $\pi=0.27\pm 0.19$~mas, 
is consistent with the spectroscopic measurement of $D_{\rm L}=3.2\pm 0.5$~kpc, 
but the uncertainty is very large due to the significant uncertainty of $\pi$.

Spectroscopic follow-up observations are important for two major scientific reasons.
First, these observations would allow one to measure the period and eccentricity 
of the planet, which has not been done before for any microlens planet.  Second, 
one can also probe for close-in planets, to which the RV method is sensitive.  We 
note that the predicted period is long, 8.6~yr for the close solution and 15.0~yr 
for the wide solution, and thus the microlensing planet can be confirmed from 
spectroscopic follow-up observations that are conducted several times per year.  
We also note that denser sampling may enable the discovery of close-in rocky planets 
and will allow the first exploration of the planetary system architecture by combining 
the RV and microlensing methods.  With these detections, the planetary system would 
be the closest analogue of the Solar System ever, with a sun-like star, a giant 
planet at the same distance as Jupiter, and close-in rocky planets.

\begin{figure}
\includegraphics[width=\columnwidth]{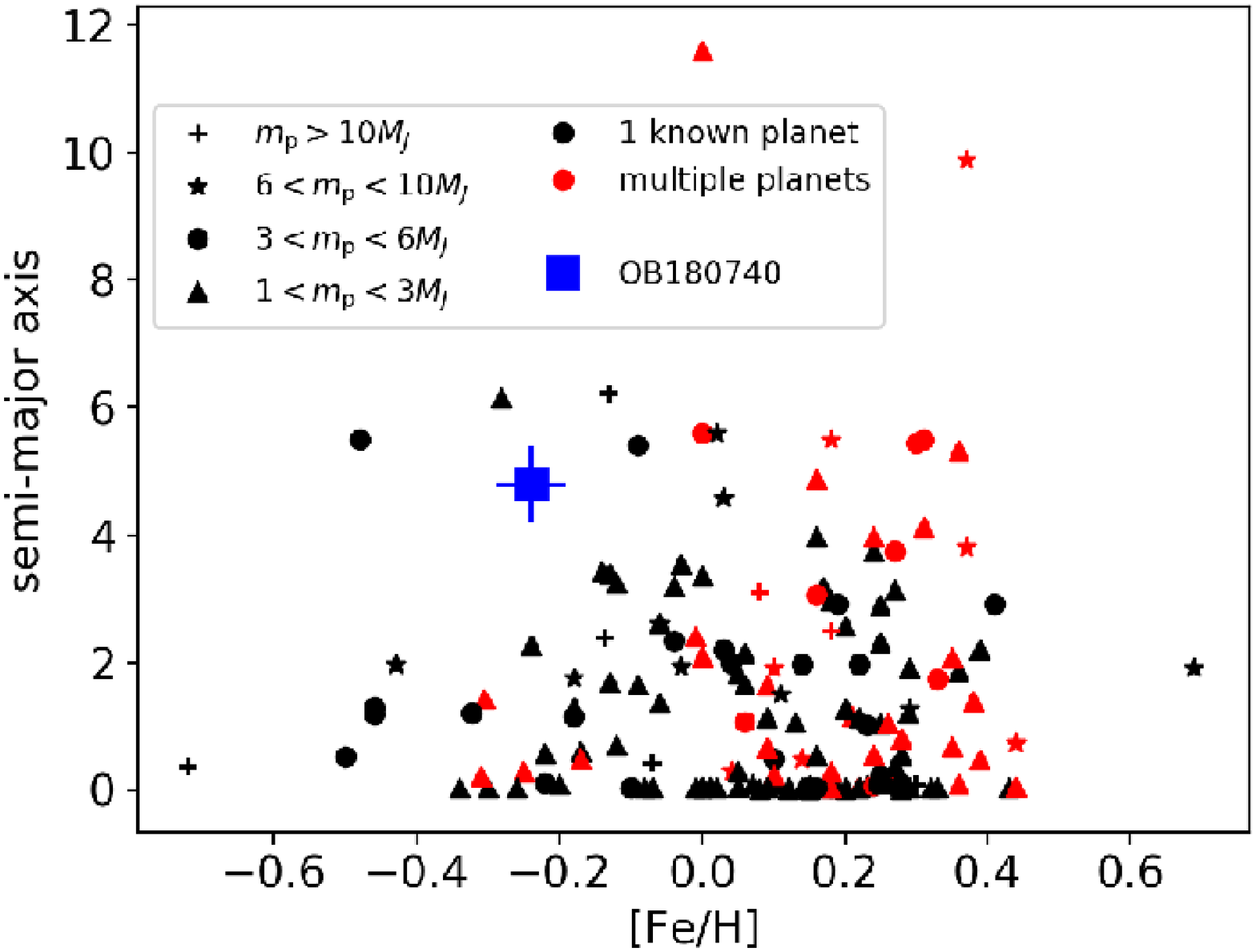}
\caption{
Comparison of OGLE-2018-BLG-0740Lb with the host star metallicities 
and semi-major axes of other exoplanets orbiting solar-type stars. Exoplanet 
data from the NASA Exoplanet Archive (downloaded 11 April 2019).
\bigskip
}
\label{fig:eleven}
\end{figure}


Considering that the blend is very likely to be the host of the planet,
the estimated metallicity of the host star of ${\rm [Fe/H]} = -0.24$ 
(section~\ref{sec:four}) indicates that planet is a super-Jupiter orbiting a 
metal-poor host star. While there is a well-known correlation between giant 
planet frequency and host star metallicity, about 3\% of metal-poor stars host 
giant planets \citep{Gonzalez1997,Santos2004,Fischer2005}.  Figure \ref{fig:eleven} 
compares OGLE-2018-BLG-0740Lb with other known giant exoplanets 
($1 M_{\rm J} \le M_{\rm p} \le 13 M_{\rm J}$) orbiting solar-type stars 
($0.8 M_\odot \le M \le 1.1 M_\odot$). The discovered microlensing planet is 
located in a relatively underpopulated portion of exoplanet parameter space 
(large, distant planets orbiting low-metallicity stars), but is similar to 
planets reported in \citet{Santos2010}, \citet{Marmier2013}, and \citet{Teske2016}, 
which are all in systems with only one known planet.

\section{Conclusion}\label{sec:eight}

We presented the analysis of the microlensing event OGLE-2018-BLG-0740,
which exhibited a strong short-term anomaly in the lensing light curve.
We tested various interpretations of the anomaly and found that the event 
was produced by a planetary system.  Despite the very strong signal, however, 
interpreting the anomaly suffered from two types of degeneracies, in which one was 
caused by the previously known close/wide degeneracy, while the other degeneracy 
was caused by the ambiguity in the normalized source radius, finite/point-source 
degeneracy, due to the incomplete coverage of the anomaly.  With the external 
information obtained from astrometric and spectroscopic observations, we identified 
that the lens was the blend and this led to the resolution of the finite/point-source 
degeneracy in strong favor of the point-source solution.  It was found that the lens 
was a planetary system composed of a super-Jupiter planet around a solar-mass star 
located at  a distance of $\sim 3$~kpc.  The bright nature of the lens combined with 
its dominance of the observed flux suggested that the period and eccentricity of the 
microlensing planet could be measured for the first time via RV observations using 
high-resolution spectrometers mounted on large telescopes.  We presented the expected 
RV amplitude for future spectroscopic observation.

\acknowledgments
Work by CH was supported by the grant (2017R1A4A1015178) of National Research 
Foundation of Korea.
Work by AG was supported by US NSF grant AST-1516842.
Work by IGS and AG were supported by JPL grant 1500811.
AG received support from the
European Research Council under the European Union's
Seventh Framework Programme (FP 7) ERC Grant Agreement n.~[321035].
This research has made use of the NASA Exoplanet Archive, which is operated by 
the California Institute of Technology, under contract with the National Aeronautics 
and Space Administration under the Exoplanet Exploration Program. 
SD acknowledges Project 11573003 supported by National Science Foundation of China (NSFC).
The MOA project is supported by JSPS KAKENHI Grant Number JSPS24253004,
JSPS26247023, JSPS23340064, JSPS15H00781, JP16H06287, and 
JP17H02871.
YM acknowledges the support by the grant JP14002006.
DPB, AB, and CR were supported by NASA through grant NASA-80NSSC18K0274. 
The work by CR was supported by an appointment to the NASA Postdoctoral 
Program at the Goddard Space Flight Center, administered by USRA through 
a contract with NASA. NJR is a Royal Society of New Zealand Rutherford 
Discovery Fellow.
The OGLE project has received funding from the National Science Centre, Poland, grant
MAESTRO 2014/14/A/ST9/00121 to AU.
This research has made use of the KMTNet system operated by the Korea
Astronomy and Space Science Institute (KASI) and the data were obtained at
three host sites of CTIO in Chile, SAAO in South Africa, and SSO in
Australia.
We acknowledge the high-speed internet service (KREONET)
provided by Korea Institute of Science and Technology Information (KISTI).
We acknowledge 
the spectral fitting done by Yang Huang and Huawei Zhang.

\end{document}